\documentclass[aps,prl,amsmath,reprint,floatfix,superscriptaddress]{revtex4-1}
\usepackage{microtype} 
\usepackage{amssymb}
\usepackage{graphicx}
\usepackage{enumerate}
\usepackage{bm}
\usepackage[version=4]{mhchem}
\usepackage{siunitx}

\RequirePackage[
  hyperindex,colorlinks,bookmarksnumbered,
  plainpages=true,pdfstartview=FitH]{hyperref}
\hypersetup{linkcolor=blue,urlcolor=blue,citecolor=blue} 
\usepackage{hyperref}
\usepackage[all]{hypcap}

\renewcommand{\vec}[1]{\bm{#1}}

\def\SRO/{\ce{Sr2RuO4}}
\def\TFL/{$T_{\mathrm{FL}}$}
\def\TFLK/{$T_{\mathrm{FL}} \!\approx\! \SI{25}{K}$}
\def\K/{$\SI{25}{K}$}
\newcommand{\TKorb}{T_{\mathrm{orb}}}
\newcommand{\TKsp}{T_{\mathrm{sp}}}

\begin{document} 

\title{Strongly Correlated Materials from a Numerical Renormalization Group Perspective:\\ 
How the Fermi-Liquid State of Sr$_2$RuO$_4$ Emerges}
\author{Fabian B.~Kugler}
\affiliation{Arnold Sommerfeld Center for Theoretical Physics, 
Center for NanoScience,\looseness=-1\,  and 
Munich Center for \\ Quantum Science and Technology,\looseness=-2\, 
Ludwig-Maximilians-Universit\"at M\"unchen, 80333 Munich, Germany}
\author{Manuel Zingl}
\affiliation{Center for Computational Quantum Physics, Flatiron Institute, 162 5th Avenue, New York, NY 10010, USA}
\author{Hugo U.~R.~Strand}
\affiliation{Center for Computational Quantum Physics, Flatiron Institute, 162 5th Avenue, New York, NY 10010, USA}
\author{Seung-Sup B.~Lee}
\affiliation{Arnold Sommerfeld Center for Theoretical Physics, 
Center for NanoScience,\looseness=-1\,  and 
Munich Center for \\ Quantum Science and Technology,\looseness=-2\, 
Ludwig-Maximilians-Universit\"at M\"unchen, 80333 Munich, Germany}
\author{Jan von Delft}
\affiliation{Arnold Sommerfeld Center for Theoretical Physics, 
Center for NanoScience,\looseness=-1\,  and 
Munich Center for \\ Quantum Science and Technology,\looseness=-2\, 
Ludwig-Maximilians-Universit\"at M\"unchen, 80333 Munich, Germany}
\author{Antoine Georges}
\affiliation{Coll\`{e}ge de France, 11 place Marcelin Berthelot, 75005 Paris, France}
\affiliation{Center for Computational Quantum Physics, Flatiron Institute, 162 5th Avenue, New York, NY 10010, USA}
\affiliation{Centre de Physique Th\'eorique, CNRS, Ecole Polytechnique, IP Paris, 91128 Palaiseau, France}
\affiliation{Department of Quantum Matter Physics, University of Geneva, 1211 Geneva 4, Switzerland}

\date{January 2, 2020}

\begin{abstract}
The crossover from fluctuating atomic constituents to a collective state 
as one lowers temperature or energy is at the heart of the 
dynamical mean-field theory description of the solid state. 
We demonstrate that the numerical renormalization group
is a viable tool to monitor this crossover in a real-materials setting.
The renormalization group flow from high to arbitrarily small energy scales
clearly reveals the emergence of the Fermi-liquid state of Sr$_2$RuO$_4$.
We find a two-stage screening process,
where orbital fluctuations are screened at much higher energies than spin fluctuations,
and Fermi-liquid behavior, concomitant with spin coherence, below a temperature of 25 K.
By computing real-frequency correlation functions, we directly observe this spin--orbital scale separation 
and show that the van Hove singularity drives strong orbital differentiation. 
We extract quasiparticle interaction parameters from the low-energy spectrum
and find an effective attraction in the spin-triplet sector.
\end{abstract}

\maketitle
\textit{Introduction.---}%
%
Atoms with partially filled shells have a spectrum of many-body eigenstates with degeneracies associated with fluctuating spin and orbital moments.
For instance, the isolated ruthenium atom in the Ru$^{4+}$ configuration, subject to an octahedral crystal field, has a ninefold degenerate ground state corresponding to spin and orbital quantum numbers $S\!=\!L\!=\!1$~\cite{Sugaon:1970kx, Georges2013}. 
In materials with strong electronic correlations, these local fluctuations can be observed at high temperature and energy through, e.g., Curie--Weiss-like spin susceptibilities.
In correlated metals, these fluctuations are suppressed as one reaches low temperature and energy. In the Fermi-liquid regime, a nondegenerate collective ground state is formed, with long-lived coherent quasiparticle excitations and susceptibilities displaying Pauli behavior~\cite{Imada:1998aa}.

How the crossover from fluctuating atomic constituents to a collective state takes place is at the heart of the dynamical mean-field theory (DMFT) description of the solid state~\cite{Georges1996}. In this theory, each atom is viewed as exchanging electrons with an environment which self-consistently represents the whole solid. The gradual suppression of local fluctuations can be thought of as a self-consistent (multistage) Kondo screening process~\cite{10.1143/PTP.32.37} of both spin and orbital moments~\cite{Stadler2015, Deng2019}.

The renormalization group (RG) is the appropriate framework to describe and monitor these crossovers as a function of energy scale. 
Indeed, Wilson's numerical renormalization group (NRG)~\cite{Wilson:1975ys} 
has been a tool of choice for solving DMFT equations for lattice models with 
few orbital degrees of freedom~\cite{Bulla2008},
with the additional merit of providing real-frequency properties at any temperature.
Following a number of two-particle applications~%
\cite{Pruschke2005,Peters2010a,Peters2010b,Peters2011,Greger2013a,Greger2013b},
recently, even three-orbital studies have become 
possible~\cite{Stadler2015, Horvat2016, Horvat2017, Stadler2019, Deng2019, Kugler2019}.
Yet, all of these works operated in the model context.
We demonstrate here that NRG can be successfully applied to an actual material, 
accounting for its electronic structure in a realistic manner 
using density functional theory (DFT) and DMFT~\cite{Kotliar2006}.

The material we focus on, \SRO/, is one of the more thoroughly studied quantum materials~\cite{RevModPhys.75.657} 
and an ideal test bed for fundamental developments in quantum many-body theories.
Besides the unconventional superconducting state below $\sim\! \SI{1.5}{K}$~\cite{Maeno1994,Mackenzie2017}, also the normal,
Hund-metal state of \SRO/~\cite{Mravlje2011,deMedici2011,Georges2013,Mravlje2016,Kim2018,Deng2019}
attracts attention, due to textbook Fermi-liquid behavior below \TFLK/~\cite{PhysRevB.57.5505, PhysRevLett.76.126, Maeno1997, PhysRevLett.81.3006, RevModPhys.75.657, Mackenzie1996b, Stricker2014}
(though signatures of quasiparticles are found up to elevated temperatures of $\sim\! \SI{600}{K}$~\cite{Mravlje2011}).
However, temperatures below \TFL/ could not be reached with controlled computational methods hitherto.

In this Letter, we show that \SRO/ undergoes a two-stage Kondo screening 
process~\cite{Stadler2015, Mravlje2016, Deng2019}, where orbital fluctuations are screened well before the spin degrees of freedom.
We determine the associated Kondo temperatures to
$\TKorb \!\approx\! \SI{6000}{K}$ and $\TKsp \!\approx\! \SI{500}{K}$, respectively, 
and show that Fermi-liquid behavior emerges when spin coherence is \emph{fully} established
below a scale of \TFLK/~%
\footnote{The orbital and spin Kondo temperatures, $\TKorb$ and $\TKsp$, give the characteristic energy scale of
the corresponding screening process and are here deduced from the maxima of the respective zero-temperature real-frequency susceptibilities~\cite{Stadler2015,Stadler2019,Kugler2019}. 
Similarly, the Fermi-liquid crossover, which corresponds to the \emph{completion} of the screening process \cite{Deng2019}, is associated with a scale
as opposed to an exact number.
In this case, we do not extract a specific value but rather compare the experimentally observed
\K/~\cite{PhysRevB.57.5505, PhysRevLett.76.126, Maeno1997, PhysRevLett.81.3006, RevModPhys.75.657}
to our numerical data and demonstrate excellent agreement.}.
With NRG as impurity solver, 
the entire DMFT calculation is performed on the real-frequency 
axis~\footnote{We note that also tensor networks have been successfully 
used to carry out DFT+DMFT calculations directly on the real-frequency 
axis~\cite{Bauernfeind2017, Bauernfeind2018}.}%
\nocite{Bauernfeind2017, Bauernfeind2018},
and we can compute correlation functions at arbitrarily low energy scales and temperatures.
Hence, we are able to go beyond previous Monte Carlo--based DFT+DMFT 
studies~\cite{Mravlje2011,Zhang2016,Strand2019,Zingl2019,Mravlje2016,Stricker2014,Deng2016,Kim2018} 
and enter deep into the Fermi-liquid regime,
even down to $T \!=\! 0$~%
\footnote{Currently, also a matrix product states (MPS) based impurity solver is being used to study
Sr$_2$RuO$_4$ at $T=0$, although on the Matsubara axis~\cite{Linden2019}.}%
\nocite{Linden2019}.
This enables us to explore the counter-intuitive observation that the more itinerant (xy) 
orbital has the smaller quasiparticle weight~\cite{RevModPhys.75.657,Bergemann2003,Mravlje2011,Zhang2016,Kim2018,Sarvestani2018,Deng2016}.
We show that this effect is driven by a van Hove singularity close to the
Fermi level, as elaborated in~\cite{Mravlje2011}.

\begin{figure*}[t!]
\includegraphics[width=\textwidth]{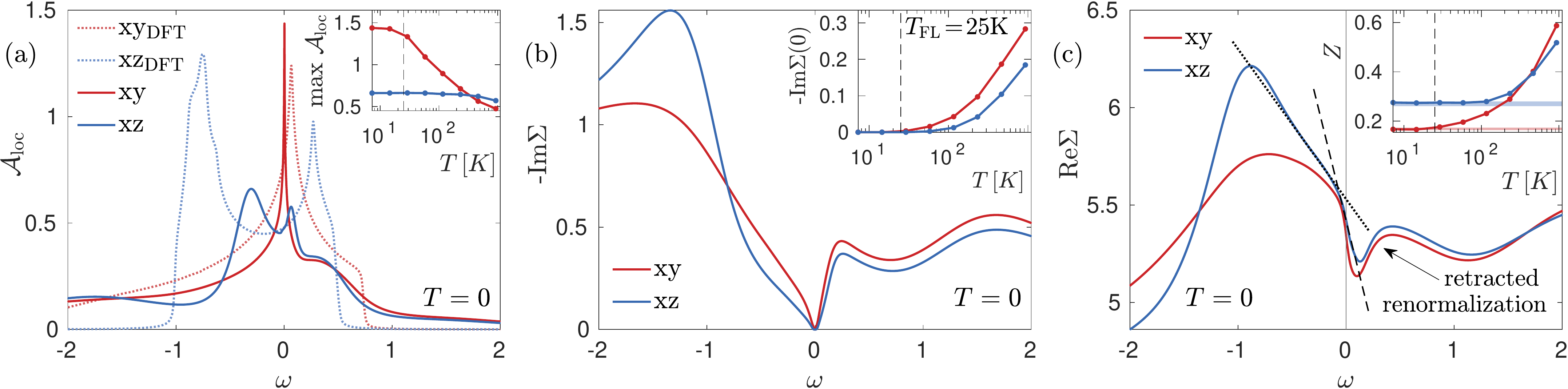}
\caption{%
Main panels: Real-frequency correlation functions at zero temperature. Insets: Characteristic values as a function of temperature, 
converging below \TFL/ (dashed line).
(a) Local spectral function, $\mathcal{A}_{\textrm{loc}}(\omega)$, from DFT+DMFT (solid lines) and DFT (dotted lines).
Inset:
$\max_\omega \mathcal{A}_{\textrm{loc}}(\omega)$.
(b) Imaginary part of the self-energy, $\textrm{Im}\Sigma(\omega)$.
Inset:
$\textrm{Im}\Sigma(\omega \!=\! 0)$.
(c) Real part of the self-energy, $\textrm{Re}\Sigma(\omega)$, 
with the two linear regimes for $\omega \!<\!0$ 
and the low-energy, positive slope for $\omega \!>\!0$ highlighted.
Inset:
$Z \!=\! (1 \!-\! \partial_\omega \mathrm{Re}\Sigma|_{\omega = 0})^{-1}$;
thick horizontal lines show
the $T \!=\! 0$ result for $Z$ calculated via renormalized parameters.
}
\label{fig:corr}
\end{figure*}

\textit{Model.---}%
%
The low-energy structure of \SRO/ can be well described by a local basis of three
maximally localized Wannier functions~\cite{MLWF1,MLWF2} with Ru-4d $t_{2g}$ symmetry denoted by 
\{xy, xz, yz\}. The corresponding non-interacting Wannier Hamiltonian is characterized by the density
of states (DOS) shown in Fig.~\ref{fig:corr}(a) below, 
reflecting the quasi-2D tetragonal crystal structure of \SRO/,
with quasi-2D xy orbitals and a strongly one-dimensional character of the degenerate xz/yz orbitals.
We employ the same Wannier Hamiltonian as in~\cite{Tamai2019, Zingl2019, Strand2019} 
(without spin-orbit coupling) combined with a local Kanamori interaction~\cite{Kanamori1963,Georges2013},
$H_{\mathrm{int}} \!=\! (U-3J)N(N-1)/2 - 2J\vec{S}^2 - J\vec{L}^2/2$,
parametrized by $U \!=\! 2.3$ and $J \!=\! 0.4$~\cite{Mravlje2011}.
Throughout this work, we use $\SI{}{eV} \!=\! 1$ as the unit of energy if not otherwise indicated.
In the Hund-metal phase of \SRO/, the pair-hopping term of the Kanamori interaction,
as part of $- J\vec{L}^2/2$,
is almost inactive.
It can thus be neglected to obtain a model with higher symmetry, 
which is more tractable for NRG, as explained in Ref.~\cite{Supp}.
\nocite{
Werner:2006rt,
Blaha2018,wien2wannier,Mostofi2014,
Weichselbaum2007,
Mitchell2014,Stadler2016,Zitko2009,Lee2016,Lee2017,
Strand:tprf}
%

\begin{figure}[t!]
\includegraphics[width=.48\textwidth]{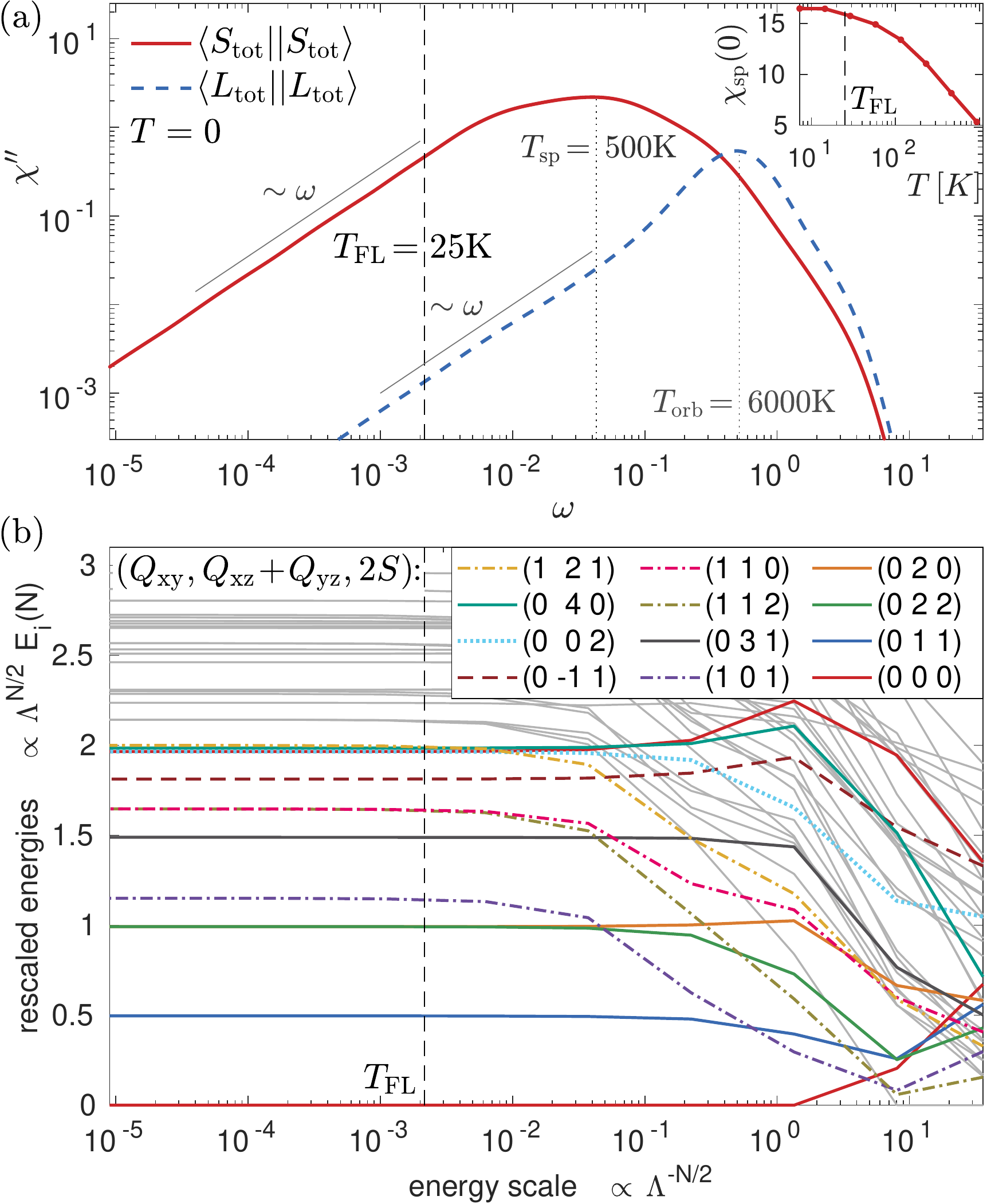}
\caption{%
(a)
Dynamic spin and orbital susceptibilities, 
$\chi^{\prime\prime}_{\mathrm{sp}}(\omega)$ and $\chi^{\prime\prime}_{\mathrm{orb}}(\omega)$, 
showing spin--orbital scale separation.
Inset: Static spin susceptibility 
as a function of temperature.
(b) NRG flow diagram, showing the rescaled eigenenergies 
(with quantum numbers given in the legend)
as a function of the energy scale \cite{Supp},
for the impurity model at self-consistency.
The spin and orbital Kondo temperatures (maximum of $\chi^{\prime\prime}$)
and the Fermi-liquid scale, \TFL/, are marked by vertical lines.%
}
\label{fig:flow}
\end{figure}

\textit{Spin--orbital separation, Fermi liquid.---}%
%
Since NRG can reach arbitrarily small energy scales,
we are able to directly observe both spin--orbital scale separation and the onset of Fermi-liquid behavior.
The zero-temperature real-frequency
orbital and spin susceptibilities~\cite{Supp}, 
$\chi^{\prime\prime}_{\mathrm{orb}}$ and 
$\chi^{\prime\prime}_{\mathrm{sp}}$, exhibit a separation of 
their maxima by more than one decade, 
see Fig.~\ref{fig:flow}(a).
This spin--orbital separation
in Kondo scales, with
$\TKorb \!\approx\! \SI{6000}{K}$ and $\TKsp \!\approx\! \SI{500}{K}$
as found from the maxima of $\chi^{\prime\prime}$,
is distinctive of correlated Hund metals~\cite{Georges2013, Stadler2015, Deng2019, Kugler2019}, 
where the Hund coupling $J$ causes 
the screening of the respective fluctuations to occur at disparate energy scales.
Further, the \emph{completed} screening of fluctuations \cite{Deng2019} is signaled by
linear behavior, 
$\chi^{\prime\prime} \!\propto\! \omega$,
found below roughly $\SI{1000}{K}$ and $\SI{25}{K}$ 
for $\chi^{\prime\prime}_{\mathrm{orb}}$ and $\chi^{\prime\prime}_{\mathrm{sp}}$, respectively.
The fully coherent Fermi-liquid state thus emerges below an energy scale of \K/.
The Fermi-liquid onset is also seen in the temperature dependence of the
static spin susceptibility, $\chi_{\mathrm{sp}}(\omega \!=\! 0)$,
which crosses over from Curie--Weiss- to Pauli-like behavior,
saturating below
\TFLK/, see inset of Fig.~\ref{fig:flow}(a).
These results clearly establish spin--orbital scale separation 
in the low-temperature Fermi-liquid state of \SRO/, 
as proposed by previous studies above \TFL/~\cite{Mravlje2016,Deng2019}.

%
A very direct observation of Fermi-liquid behavior is possible by studying the 
renormalization group flow diagram of the NRG algorithm~\cite{Bulla2008,Stadler2015,Stadler2019,Kugler2019}.
Figure~\ref{fig:flow}(b) shows the NRG Hamiltonian's (lowest) rescaled eigenenergies, 
$\Lambda^{N/2} E_i(N)$,
depending on the energy scale $\Lambda^{-N/2}$ of the RG flow, 
where $\Lambda$ is the NRG discretization parameter and 
$N$ the length of the Wilson chain \cite{Supp}.
At high energy, the states are pure atomic eigenstates, which are screened by the bath when flowing down in energy. 
Below \TFL/, the Fermi liquid is formed.
There, the flow reaches a fixed point, where the rescaled eigenenergies become independent of $N$, 
$\Lambda^{N/2} E_i(N) = E_i^*$.
The Fermi-liquid nature of this fixed point is determined by
``towers''~\cite{Bulla2008} of equidistant excitation energies 
within the same symmetry sector,
where each $E_i^*$ is composed of $n$ quasiparticle excitations,
$E_i^* = n E_{\mathrm{qp}}$.

\begin{figure}[b!]
\includegraphics[width=.409\textwidth]{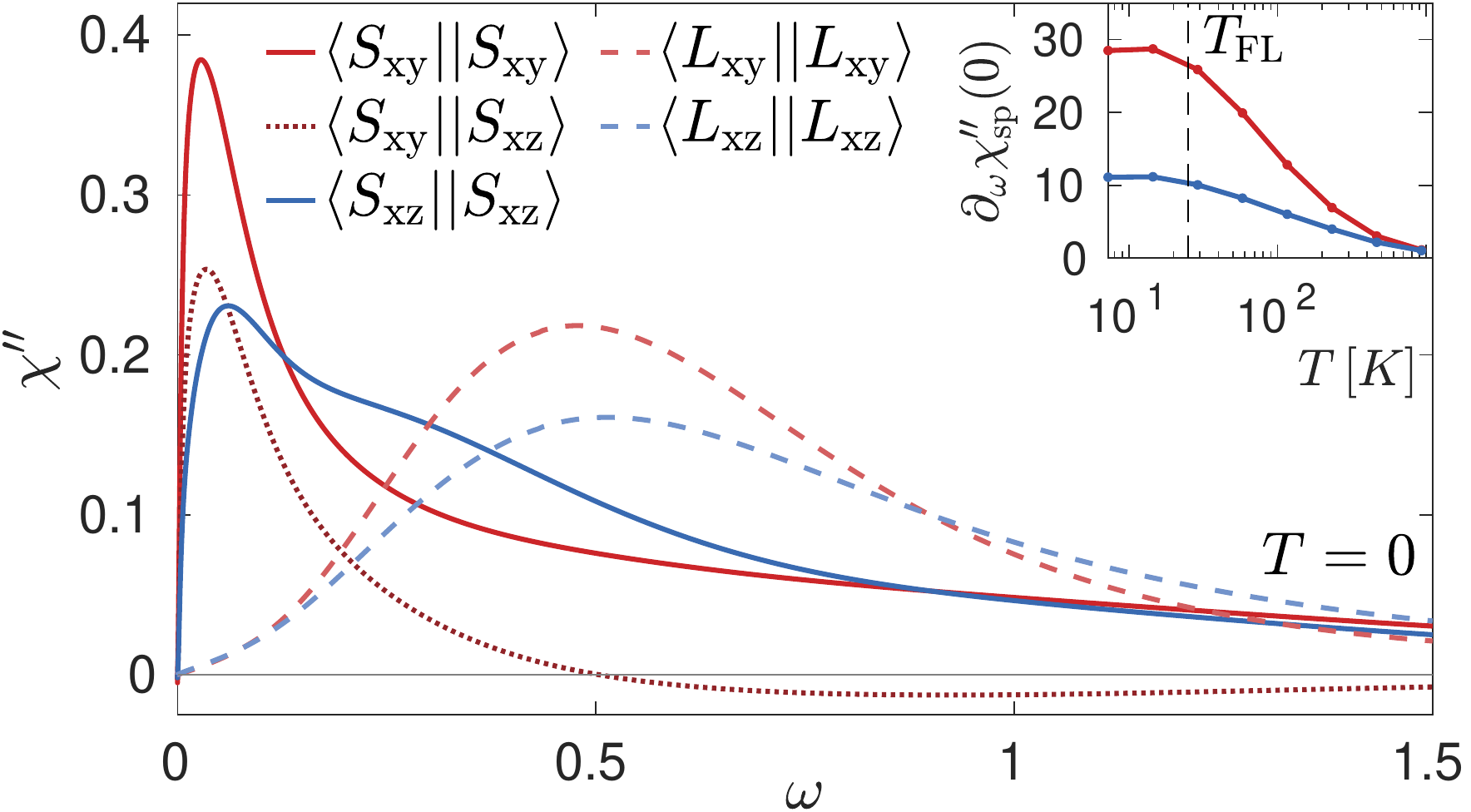}
\caption{%
Orbital-resolved, dynamic spin and angular-momentum susceptibilities,
$\chi^{\prime\prime}(\omega)$.
Inset: Temperature dependence of $\partial_\omega \chi^{\prime \prime}|_{\omega=0}$ in the spin sector, 
with \TFL/ marked as a dashed line.\vspace{1em}%
}
\label{fig:susc}
\end{figure}

Each eigenstate has the quantum numbers $(Q_{\mathrm{xy}}, Q_{\mathrm{xz}}+Q_{\mathrm{yz}}, 2S)$, 
with orbital-resolved charge $Q_m$ relative to the ground state,
and total spin, $S$.
The most prominent tower of states stems from xz/yz quasiparticles, i.e., eigenstates with quantum numbers
$(0,0,0)$, $(0,1,1)$, $(0,2,2)$, $(0,2,0)$, $(0,3,1)$, etc.; see solid lines in Fig.~\ref{fig:flow}(b). 
States with an additional xy quasiparticle are marked as dash-dotted lines.
The Fermi-liquid scale, \TFL/, is seen in the RG flow as the point where 
eigenstates with equal charge but different spin become degenerate; 
see the pairs $(0,2,0)$, $(0,2,2)$ and $(1,1,0)$, $(1,1,2)$.
Our direct evidence of the Fermi-liquid scale of \SRO/,
which conforms to the \TFLK/ found in experiments~\cite{PhysRevB.57.5505, PhysRevLett.76.126, Maeno1997, PhysRevLett.81.3006, RevModPhys.75.657},
is one of the main results of this work.

In order to understand 
how the different orbitals behave regarding
spin--orbital scale separation, 
we investigate in Fig.~\ref{fig:susc}
the orbitally resolved spin and angular-momentum susceptibilities \cite{Supp}.
We find strong 
orbital differentiation with larger amplitude in the xy than the xz spin response,
and generally a shift of spectral weight to lower frequencies in the xy compared to the xz orbital.
In nuclear magnetic resonance (NMR) spectroscopy,
the inverse nuclear spin-lattice relaxation time, $1/(T_1T)$,
is related to the zero-frequency slope of the electronic spin susceptibility, 
$1/(T_1T) \propto \partial_\omega \chi^{\prime\prime} |_{\omega=0}$ 
(neglecting matrix elements)~\cite{Alloul:2014aa, Alloul:2015aa}.
Computing the orbitally resolved $\partial_\omega \chi^{\prime\prime} |_{\omega=0}$ as a function of temperature,
we find that the xy response is about 2.5 times stronger than the xz response, 
see inset of Fig.~\ref{fig:susc}, 
in qualitative agreement with experimental~\cite{PhysRevLett.81.3006, doi:10.1143/JPSJ.67.3945, PhysRevB.64.100501} 
and theoretical works~\cite{PhysRevLett.106.096401}. 
The temperature dependence changes from linear to constant at \TFL/,
in a similar fashion for \emph{both} orbitals,
which we attribute to the strong orbital mixing on the two-particle level~\cite{Strand2019}.

%
\textit{Single-particle spectrum.---}%
%
Apart from the RG flow and (dynamic) susceptibilities,
our calculations also provide single-particle spectral information. 
Although the single-particle properties of \SRO/ have been studied 
extensively~\cite{Mravlje2011, Stricker2014, Mravlje2016, Kim2018, Zhang2016, Sarvestani2018, Tamai2019} 
using continuous-time quantum Monte Carlo (CTQMC) solvers~\cite{Gull:2011lr}, 
these calculations have a challenging scaling with inverse temperature
$\beta$, making it hard to reach the Fermi-liquid regime with 
$T \! < \! \SI{25}{K}$, i.e., $\beta \! > \! 464\,$eV$^{-1}$.
Additionally, the analytic continuation to real frequencies
severely hampers spectral resolution~\cite{Gubernatis1991}.
Here, we go beyond previous works by analyzing \SRO/ deep in the Fermi-liquid regime
at low temperatures, and even $T \!=\! 0$, directly on the real-frequency axis.

The local spectral function, $\mathcal{A}_{\textrm{loc}}(\omega)$, of \SRO/
is considerably renormalized compared to the DFT DOS~\cite{Mravlje2011, Sarvestani2018, Tamai2019}, 
see Fig.~\ref{fig:corr}(a).
When accounting for correlations, 
the spectral features are retained but shifted towards the Fermi level---%
both for the double peak in the xz/yz orbitals and the narrow xy peak.
The latter is generated by the van Hove singularity in the xy orbital, 
which is shifted towards the Fermi level by electronic correlations.
The height of the van Hove peak grows with decreasing temperature and saturates below \TFL/, 
see inset of Fig.~\ref{fig:corr}(a).

The imaginary part of the self-energy, $\textrm{Im}\Sigma(\omega)$,
shown in Fig.~\ref{fig:corr}(b),
determines the lifetime of excitations.
It has larger values at negative 
compared to positive frequencies, 
yielding shorter lifetimes for hole excitations.
Fermi-liquid behavior only emerges at frequencies below \TFL/.
The real part of the self-energy, $\textrm{Re}\Sigma(\omega)$,
displays linear (Fermi-liquid) behavior on the same small energy scale,
see Fig.~\ref{fig:corr}(c).
However, at $\omega \!\approx\! \SI{-100}{meV}$, 
it exhibits a ``kink'' leading to a second linear regime [lines in Fig.~\ref{fig:corr}(c)],
while, for $\omega$ in the range $+200\,$--$\,\SI{400}{meV}$, 
the slope of $\textrm{Re}\Sigma(\omega)$ changes sign, 
``retracting'' the renormalization of the quasiparticle dispersion. 
Hence, in this energy range, the quasiparticle velocity is larger than the bare one~\cite{Stricker2014}, as 
opposed to the usual low-energy reduction due to strong correlations.
These single-particle properties are in qualitative agreement with previous 
Monte Carlo results~\cite{Mravlje2011, Stricker2014, Mravlje2016, Kim2018, Zhang2016, Sarvestani2018, Tamai2019}.

The pronounced differentiation between the different orbitals, seen in
Figs.~\ref{fig:flow}(b) and~\ref{fig:susc},
is also reflected in the self-energy.
The xy orbital shows much stronger correlations than the xz/yz ones, 
with higher curvature in $\mathrm{Im}\Sigma(\omega)$ and 
steeper slope in $\mathrm{Re}\Sigma(\omega)$ at $\omega \!=\! 0$, 
as visible in Figs.~\ref{fig:corr}(b) and~\ref{fig:corr}(c), respectively.
The slope is related to the quasiparticle weight,
$Z \!=\! (1 \!-\! \partial_\omega \mathrm{Re}\Sigma |_{\omega = 0})^{-1}$, 
shown in the inset of Fig.~\ref{fig:corr}(c).
The zero-temperature values of $Z$ agree with 
renormalized parameters extracted directly from the spectrum
(horizontal lines, see discussion below)
and are also consistent with experiments~\cite{RevModPhys.75.657, Tamai2019}.
The low-temperature relation $Z_{\mathrm{xy}} \!<\! Z_{\mathrm{xz}}$ contrasts with
$Z_{\mathrm{xy}} \!>\! Z_{\mathrm{xz}}$ at high temperature. 
Indeed, when lowering temperature, the quasiparticle weights cross at $\sim\! \SI{350}{K}$,
and, while $Z_{\mathrm{xz}}$ levels off at $T \!\sim\! \SI{100}{K}$, $Z_{\mathrm{xy}}$ only saturates below \TFL/.
This shows that the coherence-to-incoherence crossover
and the corresponding coherence scales in \SRO/
are strongly orbital dependent~\cite{Mravlje2011,Zingl2019}. 
It is only below \TFL/ that \emph{all} orbitals are in the coherent Fermi-liquid regime.

At first sight, the stronger correlation in the xy orbital as compared to the xz/yz orbitals, 
indicated by 
$Z_{\mathrm{xy}} \!<\! Z_{\mathrm{xz}}$, is rather counterintuitive.
Usually, the ratio between the local Hubbard interaction $U$
and the bandwidth $W$, $U/W$, is a good estimator for the strength of correlations.
However, this clearly does not hold for \SRO/, since the xy orbital has a significantly larger bandwidth,
$W_{\mathrm{xy}} \!>\! W_{\mathrm{xz}}$, see Fig.~\ref{fig:corr}(a).
In~\cite{Mravlje2011}, 
it has been argued that the strong xy correlations result from
the proximity of its van Hove singularity to the Fermi level, 
see Fig.~\ref{fig:corr}(a).

\begin{figure}
\includegraphics[width=0.48\textwidth]{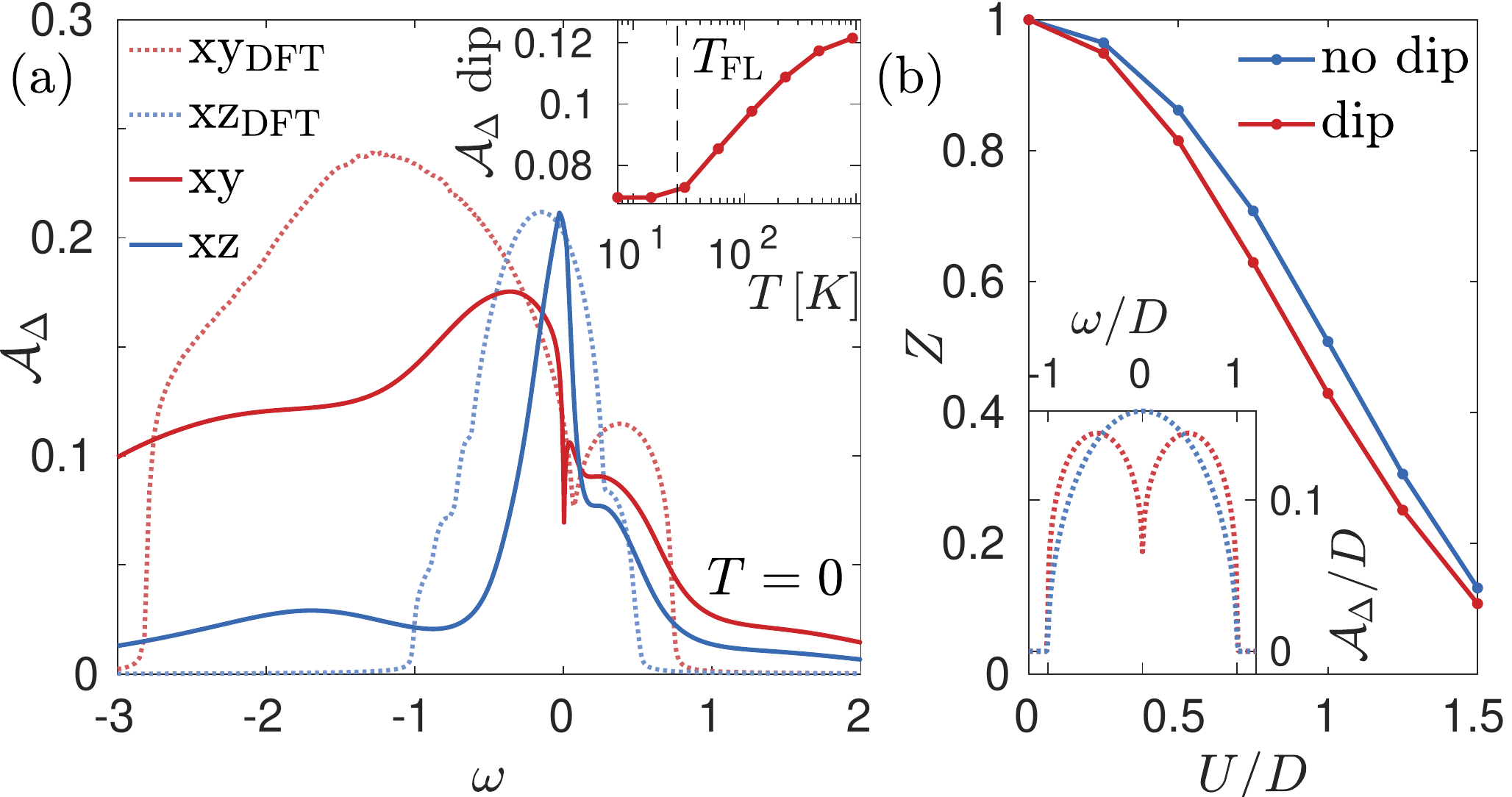}
\caption{%
(a)
Spectral function of the hybridization, $\mathcal{A}_{\Delta}(\omega)$,
in DFT+DMFT (solid lines) and DFT (dotted lines).
Inset: temperature dependence of the van Hove dip in $\mathcal{A}_{\Delta,\mathrm{xy}}(\omega)$.
(b)
Quasiparticle weight, $Z$, 
as a function of Hubbard $U$,
for a simple two-orbital model with 
identical half bandwidth, $D$.
The hybridization functions at $U \!=\! 0$ 
are shown in the inset.%
}
\label{fig:hyb}
\end{figure}

To understand this, we consider the spectral part of the
hybridization function, $\mathcal{A}_\Delta(\omega)$, of the self-consistent impurity model. 
The van Hove singularity in $\mathcal{A}_{\mathrm{loc},\mathrm{xy}}(\omega)$ 
generates a dip in $\mathcal{A}_{\Delta,\mathrm{xy}}(\omega)$ \cite{Supp}
close to zero frequency, see Fig.~\ref{fig:hyb}(a),
which implies a reduction of the effective coupling between impurity and bath at low energies for the xy orbital. 
The weaker coupling, in turn, increases the correlations and reduces the quasiparticle weight.
The temperature dependence of the dip, inset of Fig.~\ref{fig:hyb}(a), 
matches the one of $\mathrm{max}_\omega \mathcal{A}_{\mathrm{loc},\mathrm{xy}}$
in the inset of Fig.~\ref{fig:corr}(a).

To disentangle the effect of the van Hove singularity from other factors, 
we consider a simple, half-filled two-orbital model with both orbitals having the same bandwidth.
We choose a semicircular lattice DOS for one orbital and 
set the second one such that its hybridization function has a dip at zero energy, see Fig.~\ref{fig:hyb}(b).
Even in this simplified model, 
we find that $Z$ is smaller in the orbital with a dip in the hybridization. 
This suggests that the relevant measure of the correlation strength is the Hubbard interaction 
divided by the effective low-energy hybridization strength, $U/\mathcal{A}_\Delta(\omega \!=\! 0)$, 
rather than $U/W$.

\textit{Quasiparticle parameters.---}%
%
Within the NRG framework, we can extract information about the Fermi liquid and its quasiparticles
not only from correlation functions but directly from the RG flow.
To this end, we compute (zero-temperature) renormalized parameters 
from the low-energy spectrum of the (self-consistent) 
impurity model~\cite{Hewson2004,Bauer2007,Nishikawa2010, Ref2Krien2019}\nocite{Krien2019}. 
The impurity Green's function has the low-energy expansion
\begin{equation*}
G(\omega) \approx Z \cdot \big(\omega - \tilde{\epsilon} - Z \Delta(\omega) \big)^{-1} , \quad
\tilde{\epsilon} = Z \cdot \big( \epsilon + \Sigma(0) \big) .
\end{equation*}
For a (finite) Wilson chain of length $N$,
$G(\omega)$ has first-order poles at the single-particle excitation energies.
Taking the lowest particle- and hole-excitation energy $E_i(N)$ from Fig.~\ref{fig:flow}(b),
we have two equations that can be solved for 
$Z$ and $\tilde{\epsilon}$ and converged in $N$ \cite{Hewson2004}.
The results for $Z$ [and $\tilde{\epsilon}$ or $\Sigma(0)$]
are reported in \cite{Supp} and
agree quantitatively with those taken from $\Sigma(\omega)$, see inset of Fig.~\ref{fig:corr}(c).

To go beyond the single-particle picture,
we exploit that, at any finite $N$, 
there are residual quasiparticle interactions in the form of
exponentially small corrections to the equidistant tower of quasiparticle excitations.
By comparing two-particle-excitation energies $E_{mm'}^S$,
with orbital indices $m$ and $m'$ and spin index $S$,
to two single-particle excitations $E_m$ and $E_{m'}$,
the quasiparticle interaction $\tilde{U}_{mm'}^S$ is given by~\cite{Hewson2004}
\begin{equation*}
E_{mm'}^S - E_m - E_{m'} = \tilde{U}_{m m'}^S \, |\psi_m(0)|^2 \, |\psi_{m'}(0)|^2 ,
\end{equation*}
where $|\psi_m(0)|^2$ is the quasiparticle density at the impurity~\cite{Supp}.
Hence, we are in the unique position to compute quasiparticle interactions $\tilde{U}_{mm'}^S$
as well as the zero-energy real-frequency vertex $\Gamma$,  
related via $\tilde{U}_{mm'}^S = Z_m Z_{m'} \Gamma^S_{mm'}$~\cite{Hewson2004,Nishikawa2010},
for \SRO/. 
The results, listed in~\cite{Supp}, show that the orbital dependence of $\tilde{U}^S_{mm'}$ is governed by $Z_m$,
while $\Gamma^S_{mm'}$ displays only weak orbital dependence. 
Strikingly, the effective interaction in the spin-triplet sector is \emph{attractive}.
We attribute 
this
to the same mechanism as the Hund-metal $s$-wave spin-triplet superconducting instability 
found in model studies~\cite{Hoshino2015, Hoshino:2016aa}.

\textit{Conclusion.---}%
%
By following the NRG flow starting from high and proceeding to the lowest temperature and energy scales,
we have analyzed 
spin--orbital scale separation and the emergence of the Fermi liquid 
in \SRO/ within a real-materials DFT+DMFT setting.
Through linear frequency behavior of zero-temperature dynamic susceptibilities
and fixed-point analysis of the NRG flow,
we provide theoretical evidence for a Fermi-liquid scale,
in remarkable agreement with the experimentally observed
\TFLK/~\cite{PhysRevB.57.5505, PhysRevLett.76.126, Maeno1997, PhysRevLett.81.3006, RevModPhys.75.657}.
Characteristic quantities, like $\chi_{\mathrm{sp}}$ and $Z$, 
are found to converge below \K/.
Further, our real-frequency and zero-temperature results substantiate a number of features, 
such as strongly shifted spectral peaks and the peculiar frequency dependence of the self-energy, 
previously found from analytically continued Monte Carlo data \cite{Mravlje2011,Zhang2016,Strand2019,Zingl2019,Mravlje2016,Stricker2014,Deng2016,Kim2018}.
We showed that the proximity of the van Hove singularity to the Fermi level drives strong orbital differentiation in \SRO/.
Notably, the effect of van Hove singularities on the correlated state is of importance even in 
non-transition metal systems like twisted bilayer graphene~\cite{Li2009, Kerelsky2019, Choi2019}.
Finally, the extracted quasiparticle interactions $\tilde{U}^S_{mm'}$ reveal attractive coupling in the spin-triplet sector
within our \textit{ab initio} analysis. 
This paves the way towards a complete description of quasiparticles and their interactions 
in \SRO/, which are of crucial importance for the understanding of the still puzzling superconducting 
state~\cite{Mackenzie2017,Pustogow2019}.

Generally, our work demonstrates the potential of DFT+DMFT+NRG as
a new computational paradigm for real-material systems to
(i) directly access real-frequency properties at arbitrarily low temperatures and
(ii) reveal and elucidate the intricate renormalization process that occurs
during the dressing of atomic excitations by their solid-state environment.

~\\
We thank G.~Kotliar, J.~Mravlje, and A.~Weichselbaum for fruitful discussions. 
The NRG results were obtained using the QSpace tensor library~\cite{Weichselbaum2012,Weichselbaum2012a},
and TRIQS applications~\cite{Parcollet2015398, Seth2016274, TRIQS/DFTTOOLS}
were used; see~\cite{Supp} for details.
FBK, S-SBL, and JvD are supported by the Deutsche Forschungsgemeinschaft under 
Germany's Excellence Strategy--EXC-2111--390814868; 
S-SBL further by Grant.\ No.\ LE3883/2-1.
FBK acknowledges funding from the research school IMPRS-QST
and is grateful for hospitality at the Flatiron Institute, where most of this work was carried out.
The Flatiron Institute is a division of the Simons Foundation.
\bibliographystyle{apsrev4-1}
\bibliography{references}
%


\clearpage

\setcounter{page}{1}
\thispagestyle{empty}

\onecolumngrid

\begin{center}
\vspace{0.1cm}
{\bfseries\large Supplemental Material for \\
``Strongly Correlated Materials from a Numerical Renormalization Group Perspective:\\ 
How The Fermi-Liquid State of Sr$_2$RuO$_4$ Emerges''}\\
\vspace{0.4cm}
Fabian B.\ Kugler,$^1$
Manuel Zingl,$^2$
Hugo U.~R.~Strand,$^2$
Seung-Sup B.\ Lee,$^1$
Jan von Delft,$^1$
and Antoine Georges,$^{3,2,4,5}$
\\
\vspace{0.1cm}
{\it 
$^1$Arnold Sommerfeld Center for Theoretical Physics, 
Center for NanoScience,\looseness=-1\,  and 
Munich Center for \\ Quantum Science and Technology,\looseness=-2\, 
Ludwig-Maximilians-Universit\"at M\"unchen, 80333 Munich, Germany\\
$^2$Center for Computational Quantum Physics, Flatiron Institute, 162 5th Avenue, New York, NY 10010, USA \\
$^3$Coll\`{e}ge de France, 11 place Marcelin Berthelot, 75005 Paris, France \\
$^4$Centre de Physique Th\'eorique, CNRS, Ecole Polytechnique, IP Paris, 91128 Palaiseau, France\\
$^5$Department of Quantum Matter Physics, University of Geneva, 1211 Geneva 4, Switzerland
} \\
\vspace{0.6cm}
\end{center}

\twocolumngrid

In this Supplemental Material,
we first provide the definitions of the susceptibilities shown in the main text,
discuss the Hamiltonians and the neglect of pair hopping,
and give some algorithmic details.
Next, we list the quasiparticle parameters deduced from the NRG flow.
Finally, we benchmark our NRG results at various temperatures 
against continuous-time Quantum Monte Carlo (CTQMC)~\cite{Gull:2011lr} data 
obtained in the hybridization expansion (CTHYB)~\cite{Werner:2006rt, Gull:2011lr, Seth2016274}.
Citations refer to the list of references given in the main text.
\section{Susceptibilities}
We compute susceptibilities as retarded two-point correlation functions 
of bosonic operators $A$, $B$ on the impurity,
$\chi(t) = \langle A || B \rangle (t) = -i\Theta(t) \langle [A(t), B] \rangle$.
Focusing on their spectral density, we have $\chi = \chi^{\prime} - i\pi \chi^{\prime\prime}$
and $\chi^{\prime\prime}(\omega) = \tfrac{1}{2\pi} \int \mathrm{d}t \, e^{i\omega t} \langle [A(t), B] \rangle$.
Spin susceptibilities are computed via the spin operator
in $z$ direction,
\begin{equation*}
S_m = \frac{1}{2} \sum_{\sigma \sigma'} d^\dag_{m\sigma} \tau^{z}_{\sigma \sigma'} d_{m\sigma'}
,
\end{equation*}
using the Pauli matrix $\tau^z = \mathrm{diag}(1,-1)$
and the creation operator $d_{m\sigma}^{\dag}$ of an electron in orbital $m$ with spin $\sigma$ on the 
impurity.
Orbital (or angular-momentum) susceptibilities are computed via
\begin{equation*}
L_m = \frac{i}{\sqrt{2}} \sum_{\sigma m' m''} \epsilon_{m m' m''} d^\dag_{m' \sigma} d_{m'' \sigma}
\end{equation*}
with the Levi-Civita symbol $\epsilon_{mm'm''}$.
The factor of $1/\sqrt{2}$ is chosen for convenience, such that 
$\chi^{\prime\prime}_{\mathrm{sp}}$
and
$\chi^{\prime\prime}_{\mathrm{orb}}$
have roughly the same integral weight
$\int_0^{\infty} \chi^{\prime\prime} (\omega) \, \mathrm{d} \omega$.
Total susceptibilities are obtained from 
$S_{\mathrm{tot}}=\sum_m S_m$ and
$L_{\mathrm{tot}}=\sum_m L_m$.
Finally, the behavior of the orbital susceptibilities, $\langle L_m || L_m \rangle$,
is very similar to that of orbital-resolved charge susceptibilities \cite{Kugler2019},
$\langle N_m || N_m \rangle$ with
$N_m = \sum_{\sigma} d^\dag_{m\sigma} d_{m\sigma}$.

\section{Hamiltonians}
To construct the non-interacting Hamiltonian, we use maximally localized
Wannier functions~\cite{MLWF1,MLWF2} 
for the three $t_{2g}$-like orbitals centered on the Ru atoms, 
employing the software packages 
WIEN2K~\cite{Blaha2018} wien2wannier~\cite{wien2wannier}, 
wannier90~\cite{Mostofi2014} 
and TRIQS/DFTTools~\cite{Parcollet2015398, TRIQS/DFTTOOLS};
see~\cite{Tamai2019} for further details. 
The resulting Wannier Hamiltonian, $h_{mm'}(\vec{k})$, 
is nondiagonal in orbital space. However, without spin-orbital coupling, which we neglect in this work, local single-particle quantities
are orbital-diagonal due to the crystal symmetry of \SRO/.
This applies to the impurity energy levels, 
$\bm{\epsilon}_d = \sum_{\vec{k}} \bm{h}(\vec{k})$ (momentum sum normalized),
the local propagator
\begin{equation*}
\bm{G}(\omega) = \sum_{\vec{k}} [ \omega + i0^+ + \mu - \bm{h}(\vec{k}) - \bm{\Sigma}(\omega) ]^{-1}
,
\end{equation*}
and the hybridization function $\bm{\Delta}(\omega)$.

The spectral density of the hybridization is evaluated as
$\mathcal{A}_{\Delta,m}(\omega) = \tfrac{1}{\pi} \textrm{Im} [ G_m^{-1}(\omega) + \Sigma_m(\omega) ]$.
This already indicates the inverse relation between $\mathcal{A}_\Delta$ and the
spectral function $\mathcal{A}=-\tfrac{1}{\pi}\mathrm{Im}G$,
responsible for producing a dip in the hybridization from a van Hove peak in the spectrum.
Indeed, if consider small frequencies $\omega$
where $\mathrm{Im}\Sigma(\omega) \approx 0$
and assume $|\mathrm{Re}G(\omega)| \ll |\mathrm{Im}G(\omega)|$,
we directly get $\mathcal{A}_\Delta(\omega) \propto \mathcal{A}^{-1}(\omega)$.

%
\begin{table*}[t]
\begin{tabular*}{0.3\textwidth}{@{\extracolsep{\fill}} c | l l l l }
(a) & & & & 
\\
 & $Z$ & $\tilde{\epsilon}$ & $\epsilon$ & $\Sigma(0)$
\\ \hline
\rule{0pt}{2.5ex}\!
xy & $0.17$ & $-0.076$ & $-5.80$ & $5.35$
\\
xz & $0.26$ & $-0.078$ & $-5.72$ & $5.42$
\end{tabular*}%
~~~~~~~~~~
\begin{tabular*}{0.55\textwidth}{@{\extracolsep{\fill}} c | l l l l | l l}
(b) & \multicolumn{4}{c|}{$S\!=\!0$ (singlet)} & \multicolumn{2}{c}{$S\!=\!1$ (triplet)}
\\
 & xy-xy & xz-xz & xy-xz & xz-yz & xy-xz & xz-yz
\\ \hline
\rule{0pt}{2.5ex}\!
$\tilde{U}$ & $0.17$ & $0.40$ & $0.27$ & $0.40$ & $-0.085$ & $-0.12$
\\
$\Gamma$ & $6.1$ & $5.9$ & $6.1$ & $5.9$ & $-2.0$ & $-1.7$
\\
\end{tabular*}%
\ \\
\ \\
\caption{%
Zero-temperature quasiparticle parameters deduced from the NRG low-energy spectrum. 
(a) Quasiparticle weight $Z$, quasiparticle energy level $\tilde{\epsilon}$,
bare energy level $\epsilon$, and the resulting self-energy at zero frequency, $\Sigma(0)$.
(b) Quasiparticle interactions $\tilde{U}_{mm'}^S$ and the zero-energy vertex $\Gamma^S_{mm'}$,
revealing an effective attraction in the spin-triplet sector.
The two significant digits in all values give a rough estimate of the numerical accuracy.%
}
\label{tab:rpt}
\end{table*}
%

The widely used, local, SO(3)-symmetric Kanamori interaction Hamiltonian,
$H_{\mathrm{int}} = (U-3J)N(N-1)/2 - 2J\vec{S}^2 - J\vec{L}^2/2$,
consists of a density-density, spin-flip, and pair-hopping part~\cite{Georges2013},
\begin{align*}
H_{\mathrm{int}} 
& = 
H_{\mathrm{dd}} + H_{\mathrm{sf}} + H_{\mathrm{ph}} 
,
\\
H_{\mathrm{dd}} 
& = 
U \sum_m n_{m\uparrow} n_{m\downarrow}
+
U' \sum_{m \neq m'} n_{m\uparrow} n_{m'\downarrow}
\\
& \ +
(U'-J) \sum_{m < m',\sigma} n_{m\sigma} n_{m'\sigma}
,
\\
H_{\mathrm{sf}} 
& = 
- J \sum_{m \neq m'} 
d^\dag_{m\uparrow} d_{m\downarrow}
d^\dag_{m'\downarrow} d_{m'\uparrow}
,
\\
H_{\mathrm{ph}} 
& = 
J \sum_{m \neq m'} 
d^\dag_{m\uparrow} d^\dag_{m\downarrow}
d_{m'\downarrow} d_{m'\uparrow}
,
\end{align*}
where $U'=U-2J$ and shifts of the chemical potential are suppressed.
The spin-flip term is crucial for the SU(2) spin symmetry and Hund-metal physics.
By contrast, we argue in the following that the pair-hopping term is almost inactive 
in the Hund-metal phase of \SRO/
and can be neglected to obtain a model with higher symmetry that is more tractable for NRG (see below).

\section{Pair hopping}
Considering the identical prefactor $J$ of the spin-flip and pair-hopping term,
it seems \textit{a priori} hardly justified to neglect $H_{\mathrm{ph}}$.
However, it is readily understood that the effect of $H_{\mathrm{ph}}$ is a high-energy process,
requiring states with one fully occupied and one completely empty orbital at the impurity site.
At low energies, these contributions are suppressed;
the dominant contributions instead have an impurity occupation of four electrons
almost equally distributed among the three orbitals in the case of \SRO/.

Furthermore, we can \textit{a posteriori} justify neglecting $H_{\mathrm{ph}}$
by evaluating the probability for an empty and doubly occupied orbital in the 
thermal state $\rho$.
For this, we use the projectors
\begin{equation*}
P_{m \uparrow\downarrow} = n_{m\uparrow} n_{m\downarrow},
\quad
P_{m 0} = (1-n_{m\uparrow}) (1-n_{m\downarrow}),
\end{equation*}
to find that the probabilities
\begin{equation*}
\mathrm{prob}_{\mathrm{ph},m\rightarrow m'} = \mathrm{Tr} [ \, \rho \, P_{m \uparrow\downarrow} P_{m' 0} ]
\end{equation*}
are all on the level of a few percent.
We also compared imaginary-time CTHYB results with and without pair hopping
and found deviations of similar magnitude. 

\section{Algorithmic details}
Combining the quadratic part of the Hamiltonian, 
with $\Delta_{\mathrm{xy}}(\omega) \!\neq\! \Delta_{\mathrm{xz}}(\omega) \!=\! \Delta_{\mathrm{yz}}(\omega)$,
with the SO(3)-symmetric interaction Hamiltonian,
we have a \textbf{ch}arge, \textbf{orb}ital,
and \textbf{sp}in symmetry of
$\mathrm{U}(1)_{\mathrm{ch}} \otimes \mathrm{SO}(2)_{\mathrm{orb}} \otimes \mathrm{SU(2)}_{\mathrm{sp}}$.
Computationally, the one-dimensional $\mathrm{SO}(2)$ symmetry is rather weak.
However, by neglecting the pair-hopping term,
we obtain the larger symmetry
$\mathrm{U}(1)_{\mathrm{ch},\mathrm{xy}} \otimes \mathrm{U}(1)_{\mathrm{ch},\mathrm{xz}} \otimes \mathrm{U}(1)_{\mathrm{ch},\mathrm{yz}} \otimes \mathrm{SU(2)}_{\mathrm{sp}}$.

We employ the full density-matrix (fdm) NRG~\cite{Weichselbaum2007}
and exploit these symmetries using the QSpace tensor library~\cite{Weichselbaum2012,Weichselbaum2012a}.
For further efficiency, we interleave~\cite{Mitchell2014,Stadler2016} the Wilson chains of all orbitals and
thereby artificially break the symmetry between the xz and yz orbitals,
but restore it by averaging results for these orbitals at each DMFT iteration.
We use an NRG discretization parameter of $\Lambda=6$
and keep up to $10^5$ SU(2)-spin multiplets (roughly $4\cdot 10^5$ individual states)
during the iterative diagonalization.
Sufficient resolution at finite energies is obtained by
averaging results over four shifted discretization grids~\cite{Zitko2009}
and by using an adaptive broadening scheme~\cite{Lee2016,Lee2017}.

In the illustration of the NRG flow in Fig.~\ref{fig:flow}(b), 
xz and yz contributions are averaged as well.
To understand the rescaling of the axes in Fig.~\ref{fig:flow}(b), we recall that
the iterative diagonalization with a 
successively increasing Wilson chain length $N$ 
sets a characteristic energy splitting
of $a \Lambda^{-N/2}$, with $a$ of order unity \cite{Bulla2008}.
The $y$-axis is thus rescaled $\propto\! \Lambda^{-N/2}$ to have 
converged energy levels with convenient values.
Further, in fdm NRG,
temperature-dependent quantities are computed unambiguously 
by including all Wilson shells $N$
(of characteristic energy scale $E_N = a \Lambda^{-N/2}$)
with their respective, temperature-dependent weight $w_N^T$ \cite{Weichselbaum2012}.
Typically, $w_N^T$ is maximal close to $E_N \approx T$ \cite{Weichselbaum2012}.
For the $x$-axis of Fig.~\ref{fig:flow}(b),
we fix the prefactor $a$ by actually requiring 
$(\sum_N w^T_N E_N) / (\sum_N w^T_N) = T$ and thus have a unique assignment of
shell index to energy scale and temperature.

Finally, to obtain a smooth hybridization in the DMFT self-consistency iteration, 
performed entirely on the real-frequency axis,
we use a momentum summation with a large number of
$4 \cdot 10^6$ $\vec{k}$ points in the irreducible Brillouin zone
and manually set $-\mathrm{Im}\Sigma_{\mathrm{xy}}\geq 0.005$ and $-\mathrm{Im}\Sigma_{\mathrm{xz}/\mathrm{yz}}\geq 0.01$.

\begin{figure*}[t!]
\includegraphics[width=\textwidth]{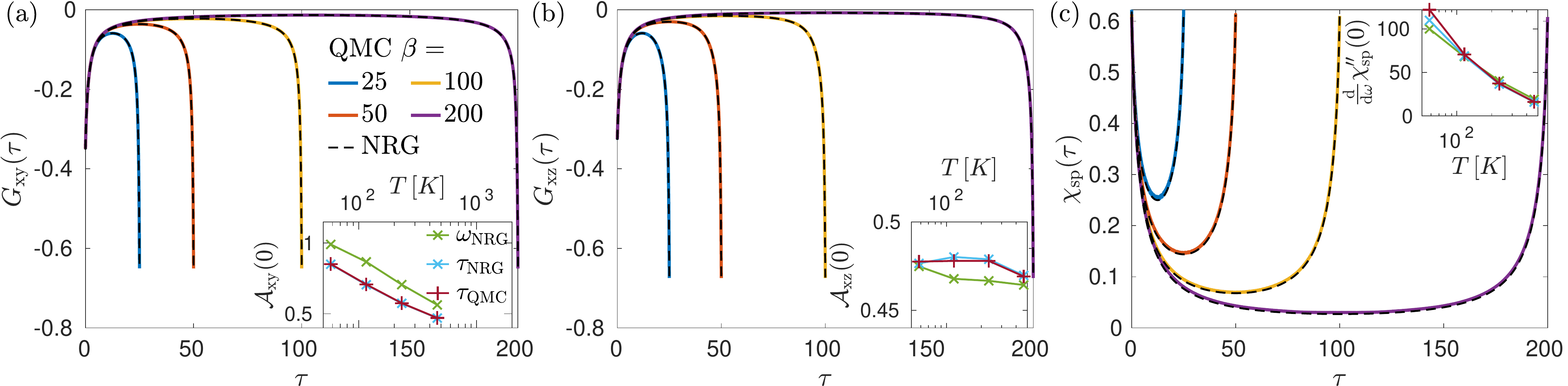}
\caption{%
Benchmark of NRG results against imaginary-time CTHYB data at various temperatures.
(a) Single-particle Green's function in the xy orbital. 
Inset: Spectral weight at zero frequency, $\mathcal{A}(0)$,
computed from the NRG real-frequency result (label $\omega_{\mathrm{NRG}}$)
and the imaginary-time proxy, $-\tfrac{\beta}{\pi}G(\tau=\beta/2)$, both 
in NRG and CTHYB (labels $\tau_{\mathrm{NRG}}$ and $\tau_{\mathrm{QMC}}$).
(b) Analogous plot for the xz orbital.
(c) Total spin susceptibility. 
Inset: Zero-frequency slope of its spectral density, $\chi_{\mathrm{sp}}^{\prime\prime}$, 
computed from the NRG real-frequency result
and the imaginary-time proxy, $-(\tfrac{\beta}{\pi})^2 \chi(\tau=\beta/2)$.%
}
\label{fig:tau}
\end{figure*}
\section{Quasiparticle parameters}
In the main text, we explained that the quasiparticle weight $Z$
and energy level $\tilde{\epsilon}$ can be extracted from the low-energy spectrum.
The results are reported in Tab.~\ref{tab:rpt}(a).
The values for $Z$ agree quantitatively with those obtained from the dynamic self-energy
via $Z = [ 1 - \partial_\omega \textrm{Re}\Sigma(0) ] ^{-1}$;
the same is true when comparing $\tilde{\epsilon} = Z \cdot \big( \epsilon + \Sigma(0) \big)$
to the zero-frequency value of the dynamic self-energy.

The results for the quasiparticle interaction $\tilde{U}_{mm'}^S$ and the zero-energy
vertex $\Gamma_{mm'}^S$ are listed in Tab.~\ref{tab:rpt}(b).
For their computation, we employed the quasiparticle density at the impurity, $|\psi_m(0)|^2$.
It is conceptually related to the excited state $| E_m \rangle$ and ground state $|\mathcal{G} \rangle$
of the renormalized impurity model according to 
$\psi_m(0) = \langle E_m | d^\dag_\sigma | \mathcal{G} \rangle$,
and practically evaluated as
\begin{align*}
|\psi_m(0)|^2
= 
\mathrm{Res}_{\omega=E_m} G_m(\omega)
= 
\frac{1}{1 - \partial_\omega \Delta(\omega) |_{\omega=E_m}}
.
\end{align*}
\section{Benchmarking NRG against QMC}
We benchmark our NRG results against CTHYB imaginary-time data at various  temperatures. 
All results are computed without pair hopping
and are converged on their respective DMFT self-consistency cycle.
We find very good agreement for the single-particle Green's function 
$G_{\mathrm{xy}}$ and $G_{\mathrm{xz}}$, see Fig.~\ref{fig:tau}(a) and (b).
The (total) spin susceptibilities show satisfactory agreement as well,
with slightly higher deviations, see Fig.~\ref{fig:tau}(c).

In the insets of Fig.~\ref{fig:tau},
we show the real-frequency quantities 
$\mathcal{A}(0)$ and $\tfrac{\mathrm{d}}{\mathrm{d}\omega}\chi^{\prime\prime} |_{\omega=0}$ and
their imaginary-time proxies, given by the l.h.s.\ of the relations
\begin{align*}
-\frac{\beta}{\pi}G(\tau \!=\! \beta/2) 
& = 
\mathcal{A}(0) + \frac{\pi^2}{2\beta^2} \frac{\mathrm{d}^2}{\mathrm{d}^2\omega}\mathcal{A}(0) + \mathit{O}(\beta^{-4})
,
\\
\Big(\frac{\beta}{\pi}\Big)^2\chi(\tau \!=\! \beta/2) 
& = 
\frac{\mathrm{d}}{\mathrm{d}\omega}\chi^{\prime\prime}(0) 
+ 
\frac{\pi^2}{3\beta^2} \frac{\mathrm{d}^3}{\mathrm{d}^3\omega}\chi^{\prime\prime}(0) 
+ 
\mathit{O}(\beta^{-4})
.
\end{align*}
By fitting the corresponding polynomials to the real-frequency curves, 
we find that the $\beta^{-2}$ corrections on the r.h.s.\ amount to roughly 10\%, 5\%, and 30\% for 
$G_{\mathrm{xy}}$, $G_{\mathrm{xz}}$, $\chi_{\mathrm{sp}}$, respectively. 
This is consistent with the notable deviations 
between the real-frequency values and their imaginary-time proxies
in the insets of Fig.~\ref{fig:tau}.
The CTHYB results have been obtained using
the TRIQS/CTHYB solver~\cite{Seth2016274}
and the TRIQS/TPRF package~\cite{Strand:tprf}, which are based
on the TRIQS library~\cite{Parcollet2015398}.

\end{document}